\documentclass[
reprint,
superscriptaddress,
showpacs,preprintnumbers,
footinbib,
bibnotes,
amsmath,amssymb,
aps,
pra,
floatfix,
]{revtex4-2}

\usepackage{graphicx}
\usepackage{color}

\def\Jac{\textsc{Jac}}

\newcommand{\AT}{$\to$}
\newcommand{\RT}{\rotatebox[origin=c]{180}{$\Lsh$}}

\begin{document}

\title{Multiple Photodetachment of Oxygen Anions via K-Shell Excitation and Ionization:\\ Direct Double-Detachment Processes and Subsequent Deexcitation Cascades}
%\thanks{A footnote to the article title}%

\author{S.~Schippers}
\email{stefan.schippers@physik.uni-giessen.de}
\affiliation{I.~Physikalisches Institut, Justus-Liebig-Universit\"{a}t Gie{\ss}en, Heinrich-Buff-Ring 16, 35392 Giessen, Germany}

\author{A.~Hamann}
\affiliation{I.~Physikalisches Institut, Justus-Liebig-Universit\"{a}t Gie{\ss}en, Heinrich-Buff-Ring 16, 35392 Giessen, Germany}

\author{A.~Perry-Sassmannshausen}
\affiliation{I.~Physikalisches Institut, Justus-Liebig-Universit\"{a}t Gie{\ss}en, Heinrich-Buff-Ring 16, 35392 Giessen, Germany}

\author{T.~Buhr}
\affiliation{I.~Physikalisches Institut, Justus-Liebig-Universit\"{a}t Gie{\ss}en, Heinrich-Buff-Ring 16, 35392 Giessen, Germany}

\author{A.~M\"{u}ller}
\affiliation{I.~Physikalisches Institut, Justus-Liebig-Universit\"{a}t Gie{\ss}en, Heinrich-Buff-Ring 16, 35392 Giessen, Germany}

\author{M.~Martins}
\affiliation{Institut f\"{u}r Experimentalphysik, Universit\"{a}t Hamburg, Luruper Chaussee 149, 22761 Hamburg, Germany}

\author{S.~Reinwardt}
\affiliation{Institut f\"{u}r Experimentalphysik, Universit\"{a}t Hamburg, Luruper Chaussee 149, 22761 Hamburg, Germany}

\author{F.~Trinter}
\affiliation{Institut f\"{u}r Kernphysik, Goethe-Universit\"{a}t Frankfurt am Main, Max-von-Laue-Stra{\ss}e 1, 60438 Frankfurt am Main, Germany}
\affiliation{Molecular Physics, Fritz-Haber-Institut der Max-Planck-Gesellschaft, Faradayweg 4-6, 14195 Berlin, Germany}

\author{S.~Fritzsche}
\affiliation{Helmholtz-Institut Jena, Fr{\"o}belstieg 3, 07743 Jena, Germany}
\affiliation{Theoretisch-Physikalisches Institut, Friedrich-Schiller-Universit\"{a}t Jena, Max-Wien-Platz 1, 07743 Jena, Germany}
\affiliation{GSI Helmholtzzentrum f\"u{}r Schwerionenforschung, Planckstra{\ss}e 1, 64291 Darmstadt, Germany}

\date{\today}% It is always \today, today,
%  but any date may be explicitly specified

\begin{abstract}
Experimental cross sections for $m$-fold photodetachment ($m=2-5$) of oxygen anions via $K$-shell excitation and ionization were measured in the photon-energy range of 525--1500~eV using the photon-ion merged-beams technique at a synchrotron light source. The measured cross sections exhibit clear signatures of direct double detachment, including double $K$-hole creation. The shapes of the double-detachment cross sections as a function of photon energy are in accord with Pattard's [J.~Phys.~B 35, L207 (2002)] empirical scaling law.  We have also followed the complex deexcitation cascades that evolve subsequently to the initial double-detachment events by systematic large-scale cascade calculations.  The resulting theoretical product charge-state distributions are in good agreement with the experimental findings.
\end{abstract}
	
\maketitle

\section{Introduction}

Atomic anions are peculiar quantum systems since their extra electron  is (weakly) bound by short-range forces resulting from the induced polarization of the atomic electron shells {\color{black} \cite{Hotop1975,Pegg2004b,Andersen2004b}.   In the O$^-$($1s^2\,2s^2\,2p^5\;^2P_{3/2}$) ground level, the binding energy of the most loosely bound electron is only 1.461~eV \cite{Blondel2001a}, while the ionization energy of neutral oxygen atoms amounts to 13.681~eV~\cite{Kramida2021}}. A fundamental process that allows one  to probe  electron-electron correlation effects in atomic anions is single photodetachment (SD), i.e., the removal  of an  electron by absorption of a single photon. Recent studies on multiple photodetachment of F$^-$  \cite{Mueller2018b} and C$^-$  \cite{Perry-Sassmannshausen2020}  have somewhat unexpectedly revealed that atomic anions are also ideal systems for studying  direct double detachment (DD), where one photon simultaneously ejects two electrons (see also~\cite{Schippers2020a}). For F$^-$, the cross section for simultaneous ejection of a $1s$ and a $2p$ electron could be extracted from the measured data for photon energies that range from below the threshold for direct double detachment of a  $1s$ and a $2p$ electron ($1s$+$2p$~DD) to well beyond the cross-section maximum.   This result has already stimulated a successful effort to theoretically describe the cross section for this many-particle effect with two electrons ending up in the continuum \cite{Kheifets2020}. For C$^-$, the cross section for double-core hole formation by a single photon ($1s$+$1s$ DD) was measured. Surprisingly, this cross section was about an order of  magnitude larger than what was expected on the basis of previous results on the double core-hole creation in carbon-containing molecules~\cite{Lablanquie2016}.

In order to shed more light on the role of DD processes in the inner-shell photodetachment of atomic anions, the present study extends the above mentioned previous work to the O$^-$ anion. Double detachment, or double ionization for neutral atoms and positive ions, has been repeatedly studied in the literature, mostly for  helium-like systems. Since a historical account has already been given in Ref.~\cite{Mueller2018b} we shall mention here only the work that is most closely related to the present investigation, i.e.,  studies on  DD of H$^-$~\cite{Donahue1982}, He$^-$~\cite{Bae1983}, and K$^-$~\cite{Bae1988} where the corresponding cross sections were scanned over very narrow photon-energy ranges of a few meV in the vicinity of the respective thresholds. Earlier studies on net double and triple {\color{black}inner-shell} detachment of O$^-$ ions also  considered only quite limited photon-energy ranges and focussed on the $1s\,2s^2\,2p^6$ resonance at 525.6~eV that is associated with $1s\to2p$ photoexcitation \cite{Gibson2012,Schippers2016a}.

When compared with the previous measurements, the present photon-energy range of 525--1500~eV is much wider and comprises the thresholds for direct single detachment of a $1s$ electron ($1s$ SD) at 529.6~eV \cite{Schippers2016a} as well as the thresholds  for $1s$+$2p$~DD, $1s$+$2s$~DD, and $1s$+$1s$~DD. The threshold for the latter process occurs at about 1100~eV (see below). Our experimental cross sections $\sigma_m$ for $m$-fold photodetachment of O$^-$ ions with $m = 2,3,4,5$ exhibit signatures of all these processes. $m$-fold photoionization  results in the production of multiply positively charged O$^{(m-1)+}$ ions and can be represented as
\begin{equation}\label{eq:reaction}
	h\nu + \textrm{O}^- \to \textrm{O}^{(m-1)+} + m\, e^-.
\end{equation}
In addition to experimental cross sections  for the production of O$^{(m-1)+}$ ions, we present theoretical results for  O$^-$ photoabsorption. Using a recently developed theoretical toolbox \cite{Fritzsche2019,Fritzsche2021},  we model the complex deexcitation cascades that follow the creation of single or double inner-shell holes and thus give rise to photon-energy-dependent distributions of product-ion charge states.

\section{Experiment}\label{sec:exp}

The measurements were carried out using  the photon-ion merged-beams technique \cite{Schippers2016} at the PIPE facility \cite{Schippers2014,Mueller2017,Schippers2020}, which is a permanently installed end-station at the photon beamline P04 \cite{Viefhaus2013} of the synchrotron radiation source PETRA\,III operated by DESY in Hamburg, Germany. Oxygen anions were produced  by a Cs-sputter ion source \cite{Middleton1984} with a sputter target made of solid aluminum-oxide and a sputter potential of about 2~kV.   After acceleration to a kinetic energy of 6~keV, the ions were passed through an analyzing dipole magnet which was adjusted such that $^{16}$O$^-$ ions were selected for further transport to the photon-ion merged-beams interaction region. The ion current in the interaction region was up to 350~nA (18 nA) with the ion beam being uncollimated (collimated to about $2\times2$~mm$^2$). These values are an order of magnitude larger than what could be achieved in our previous study where a different type of ion source was used \cite{Schippers2016a}.
	
In the interaction region, where the residual-gas pressure was in the mid $10^{-10}$~mbar range, the ion beam was coaxially merged with the counter-propagating soft x-ray photon beam over a distance of about 1.7~m. O$^{(m-1)+}$ ions  as obtained from multiple photodetachment (Eq.~\ref{eq:reaction}) were separated from the primary ion beam by a second dipole magnet. Inside this magnet, a Faraday cup collected the primary ion beam, while the charge-selected product ions were directed to the detector chamber. Along their flight path, they first passed through a spherical 180-degree out-of-plane deflector to suppress background from stray electrons, photons, and ions before the ions then entered a single-particle detector with nearly 100\% detection efficiency \cite{Rinn1982}.

The dark-count rate of our detector amounted to $\sim$0.02~Hz (see also \cite{Mueller2021b}). In each product-ion channel, this contributed to a photon-energy-independent background count rate. Other contributions to the background count rate arose from collisions of the primary ions with residual-gas particles, where electrons can be lost from the projectiles by stripping reactions. This contribution to the energy-independent background was particularly noticeable in the double-detachment channel, but practically negligible for triple, fourfold, and fivefold detachment. {\color{black}For double detachment, the background was determined by recording O$^+$ product ions while alternatingly switching the photon beam on and off by opening and closing a fast beam shutter in the photon beamline.}

Relative cross sections for $m$-fold photodetachment were obtained by normalizing the background-subtracted count rates of O$^{(m-1)+}$ product ions on the primary O$^-$ ion current and on the photon flux measured with a calibrated photodiode.  The energy-dependent photon flux reached its peak of $3\times10^{13}$~s$^{-1}$ at photon energies around 850~eV and at a photon-energy spread of $\sim$1~eV. This spread also depended on the photon energy ranging from $\sim$0.4~eV at 500~eV up to $\sim$2.5~eV at 1500~eV for a nominal width of the monochromator exit slit of 1000~$\mu$m. The relative cross sections were put on an absolute scale by using our previously measured absolute cross sections for single and double detachment of O$^-$ as reference values \cite{Schippers2016a}. The systematic uncertainty of the experimental absolute cross sections is estimated to be $\pm$15\% at 90\% confidence level \cite{Schippers2014}.

The photon-energy scale was calibrated by absorption measurements in nitrogen gas using the lowest vibrational component of the $1s\to\pi^*$ resonance at 400.86(3)~eV (see discussion by M\"uller et~al.~\cite{Mueller2018c} and references therein) as a calibration point. A second (intrinsic) calibration point was provided by the O$^-$($1s\to2p$) resonance at 525.6(1)~eV \cite{Schippers2016a}. In addition, the Doppler shift that is associated with the ion-beam velocity was taken into account. {\color{black} It amounted to 0.467~eV at 520~eV and to 1.346~eV at 1500~eV}. The remaining uncertainty of the calibrated photon-energy scale is estimated to be $\pm$0.2~eV for photon  energies of less than 600~eV.  Since there are no calibration points at higher energies, the  uncertainty increases with increasing photon energy.  At a photon energy of 1500~eV, it is estimated to amount to about~$\pm$2~eV.

\section{Computations}\label{sec:theo}

\begin{table}[b]
\caption{\label{tab:thres}Computed threshold energies for single detachment (SD) and double detachment (DD) of O$^-$. The energy ranges comprise all fine-structure levels that belong to the specified configurations. }
  \begin{ruledtabular}
\begin{tabular}{rcc}
  Configuration                        & Process             &   Energy (eV) \\    \hline
      $1s^1\,2s^2\,2p^5$          &     $1s$ SD        &      529.8--532.5  \\
     $1s^1\,2s^2\,2p^4$            &    $1s$+$2p$ DD  &    545.2--553.8  \\
     $1s^1\,2s^1\,2p^5$            &    $1s$+$2s$ DD  &      561.9--574.0   \\
     $2s^2\,2p^5$                      &     $1s+1s$ DD  &        1160.2        \\
 \end{tabular}
\end{ruledtabular}
\end{table}

\begin{table}
\caption{\label{tab:casc1s}Simplified scheme of the cascade that sets in subsequent to $1s$ SD of O$^-$($1s^2\,2s^2\,2p^5$). The scheme contains all energetically allowed single-step autoionizing transitions and dipole-allowed radiative transitions. Right arrows and down-right arrows denote Auger transitions from the configuration to the left and radiative transitions from a configuration above, respectively. The ground configurations of the different product ions,  which are usually reached via multiple pathways, are printed in bold face.}
  \begin{ruledtabular}
\begin{tabular}{ll@{}l@{}ll@{}l@{}ll@{}l@{}l}
 &\multicolumn{3}{c}{O} & \multicolumn{3}{c}{O$^{+}$} & \multicolumn{3}{c}{O$^{2+}$} \\
 \hline
    &   \multicolumn{3}{l}{$1s^1\,2s^2\,2p^5$}          &    \multicolumn{3}{l}{\AT\ $\boldsymbol{1s^2\,2s^2\,2p^3}$}      &  & & \\
     &                   & &                                                    &    \multicolumn{3}{l}{\AT\ $1s^2\,2s^1\,2p^4$}      &    \multicolumn{3}{l}{\AT\ $\boldsymbol{1s^2\,2s^2\,2p^2}$}     \\
     &                   &  &                                                   &   &  \multicolumn{2}{l}{\RT\ $\boldsymbol{1s^2\,2s^2\,2p^3}$}  &  & & \\
     &                   &   &                                                  &    \multicolumn{3}{l}{\AT\ $1s^2\,2s^0\,2p^5$}      &     \multicolumn{3}{l}{\AT\ $1s^2\,2s^1\,2p^3$}    \\
     &                    &  &                                                  &         & &                                                                    &    & \multicolumn{2}{l}{\RT\  $\boldsymbol{1s^2\,2s^2\,2p^2}$}    \\
     &                    &   &                                                 &   &  \multicolumn{2}{l}{\RT\ $1s^2\,2s^1\,2p^4$}  &     \multicolumn{3}{l}{\AT\ $\boldsymbol{1s^2\,2s^2\,2p^2}$}\\
      &                   &    &                                                &    & & \RT\ $\boldsymbol{1s^2\,2s^2\,2p^3}$                               &  & & \\
  & & \multicolumn{2}{l}{\RT\ $1s^2\,2s^2\,2p^4$}  &                            & &                                                 &  & &             \\
  & & \multicolumn{2}{l}{\RT\ $1s^2\,2s^1\,2p^5$}  &    \multicolumn{3}{l}{\AT\ $\boldsymbol{1s^2\,2s^2\,2p^3}$}      &  & &             \\
 &  & & \RT\ $\boldsymbol{1s^2\,2s^2\,2p^4}$                               &         & &                                                                    &  & &\\
  \end{tabular}
\end{ruledtabular}
\end{table}

\begin{table}
\caption{\label{tab:casc1s2p}Simplified scheme of the cascade that sets in subsequent to $1s$+$2p$ DD of O$^-$($1s^2\,2s^2\,2p^5$). The scheme contains all energetically allowed single-step autoionizing transitions and dipole-allowed radiative transitions. See Tab.~\ref{tab:casc1s} for further details.}
  \begin{ruledtabular}
\begin{tabular}{ll@{}l@{}ll@{}l@{}ll@{}l@{}l}
& \multicolumn{3}{c}{O$^+$} & \multicolumn{3}{c}{O$^{2+}$} & \multicolumn{3}{c}{O$^{3+}$} \\
 \hline
   &    \multicolumn{3}{l}{$1s^1\,2s^2\,2p^4$}          &    \multicolumn{3}{l}{\AT\ $\boldsymbol{1s^2\,2s^2\,2p^2}$}      &  & & \\
    &                    & &                                                    &    \multicolumn{3}{l}{\AT\ $1s^2\,2s^1\,2p^3$}      &    \multicolumn{3}{l}{\AT\ $\boldsymbol{1s^2\,2s^2\,2p^1}$}     \\
     &                   &  &                                                   &   &  \multicolumn{2}{l}{\RT\ $\boldsymbol{1s^2\,2s^2\,2p^2}$}  &  & & \\
    &                    &   &                                                  &    \multicolumn{3}{l}{\AT\ $1s^2\,2s^0\,2p^4$}      &     \multicolumn{3}{l}{\AT\ $1s^2\,2s^1\,2p^2$}    \\
    &                     &  &                                                  &         & &                                                                    &    & \multicolumn{2}{l}{\RT\  $\boldsymbol{1s^2\,2s^2\,2p^1}$}    \\
    &                     &   &                                                 &   &  \multicolumn{2}{l}{\RT\ $1s^2\,2s^1\,2p^3$}  &     \multicolumn{3}{l}{\AT\ $\boldsymbol{1s^2\,2s^2\,2p^1}$}\\
    &                     &    &                                                &    & & \RT\ $\boldsymbol{1s^2\,2s^2\,2p^2}$                               &  & & \\
  & & \multicolumn{2}{l}{\RT\ $1s^2\,2s^2\,2p^3$}  &                            & &                                                 &  & &             \\
  & & \multicolumn{2}{l}{\RT\ $1s^2\,2s^1\,2p^4$}  &    \multicolumn{3}{l}{\AT\ $\boldsymbol{1s^2\,2s^2\,2p^2}$}      &  & &             \\
  & & & \RT\ $\boldsymbol{1s^2\,2s^2\,2p^3}$                               &         & &                                                                    &  & &\\
  \end{tabular}
\end{ruledtabular}
\end{table}

\begin{table}
\caption{\label{tab:casc1s2s}Simplified scheme of the cascade that sets in subsequent to $1s$+$2s$ DD of O$^-$($1s^2\,2s^2\,2p^5$). The scheme contains all energetically allowed single-step autoionizing transitions and dipole-allowed radiative transitions. See Tab.~\ref{tab:casc1s} for further details. In principle, the O$^{3+}$($1s^2\,2s^1\,2p^2$) $\to$  O$^{4+}$($1s^2\,2s^2$) and O$^{3+}$($1s^2\,2s^0\,2p^3$) $\to$  O$^{4+}$($1s^2\,2s\,2p$) transitions could be conceived, but these are energetically not possible.}
  \begin{ruledtabular}
\begin{tabular}{ll@{}l@{}l@{}ll@{}l@{}ll@{}l@{}ll}
 & \multicolumn{4}{c}{O$^+$}  & \multicolumn{3}{c}{O$^{2+}$} & \multicolumn{3}{c}{O$^{3+}$}  \\
 \hline
     &  \multicolumn{4}{l}{$1s^1\,2s^1\,2p^5$}                                        &    \multicolumn{3}{l}{\AT\ $1s^1\,2s^2\,2p^3$}                            &  \multicolumn{3}{l}{\AT\ $\boldsymbol{1s^2\,2s^2\,2p^1}$}    \\
       &                 & &    &                                                                           &    & &                                                                                               &    \multicolumn{3}{l}{\AT\ $1s^2\,2s^1\,2p^2$}    \\
         &               & &    &                                                                           &    & &                                                                                               &    & \multicolumn{2}{l}{\RT\ $\boldsymbol{1s^2\,2s^2\,2p^1}$}     \\
        &                & &    &                                                                           &    & &                                                                                               &    \multicolumn{3}{l}{\AT\ $1s^2\,2s^0\,2p^3$}   \\
       &                  &  &   &                                                                          &         & &                                                                                          &    & \multicolumn{2}{l}{\RT\  $1s^2\,2s^1\,2p^2$}     \\
       &                  &  &   &                                                                          &         & &                                                                                          &    &  &  \RT\  $\boldsymbol{1s^2\,2s^2\,2p^1}$   \\
        &                &  &    &                                                                          &   &  \multicolumn{2}{l}{\RT\ $\boldsymbol{1s^2\,2s^2\,2p^2}$}  &  & &   \\
        &                & &    &                                                                           &    \multicolumn{3}{l}{\AT\ $1s^2\,2s^1\,2p^3$}                            &    \multicolumn{3}{l}{\AT\ $\boldsymbol{1s^2\,2s^2\,2p^1}$}     \\
        &                &  &    &                                                                          &   &  \multicolumn{2}{l}{\RT\ $\boldsymbol{1s^2\,2s^2\,2p^2}$}  &  & &  &\\
         &               &   &   &                                                                          &    \multicolumn{3}{l}{\AT\ $1s^2\,2s^0\,2p^4$}                             &     \multicolumn{3}{l}{\AT\ $1s^2\,2s^1\,2p^2$}      \\
          &               &  &   &                                                                          &         & &                                                                                          &    & \multicolumn{2}{l}{\RT\  $\boldsymbol{1s^2\,2s^2\,2p^1}$}     \\
          &               &   &  &                                                                          &   &  \multicolumn{2}{l}{\RT\ $1s^2\,2s^1\,2p^3$}                        &     \multicolumn{3}{l}{\AT\ $\boldsymbol{1s^2\,2s^2\,2p^1}$}  \\
           &              &    &  &                                                                          &    & & \RT\ $\boldsymbol{1s^2\,2s^2\,2p^2}$                              &  & &   \\
   & & \multicolumn{3}{l}{\RT\ $1s^1\,2s^2\,2p^4$}                                 &   \multicolumn{3}{l}{\AT\ $\boldsymbol{1s^2\,2s^2\,2p^2}$}       &  & &        \\
         &                &    &  &                                                                          &   \multicolumn{3}{l}{\AT\ $1s^2\,2s^1\,2p^3$}                               &    &  &\\
        &                 &    &  &                                                                          &    & \multicolumn{2}{l}{\RT\ $\boldsymbol{1s^2\,2s^2\,2p^2}$}   &    &  &\\
         &               &   &   &                                                                          &    \multicolumn{3}{l}{\AT\ $1s^2\,2s^0\,2p^4$}                             &     \multicolumn{3}{l}{\AT\ $1s^2\,2s^1\,2p^2$}   \\
         &                &  &   &                                                                          &         & &                                                                                          &    & \multicolumn{2}{l}{\RT\  $\boldsymbol{1s^2\,2s^2\,2p^1}$}    \\
         &                &   &  &                                                                          &   &  \multicolumn{2}{l}{\RT\ $1s^2\,2s^1\,2p^3$}                        &     \multicolumn{3}{l}{\AT\ $\boldsymbol{1s^2\,2s^2\,2p^1}$}  \\
          &               &    &  &                                                                          &    & & \RT\ $\boldsymbol{1s^2\,2s^2\,2p^2}$                              &  & &  \\
  & & \multicolumn{3}{l}{\RT\ $1s^2\,2s^1\,2p^4$}                                  &    \multicolumn{3}{l}{\AT\ $\boldsymbol{1s^2\,2s^2\,2p^2}$}      &  & &        \\
  & & & \multicolumn{2}{l}{\RT\ $\boldsymbol{1s^2\,2s^2\,2p^3}$}        &                             & &                                                                      &  & &\\
  & & \multicolumn{3}{l}{\RT\ $1s^2\,2s^0\,2p^5$}                                  &    \multicolumn{3}{l}{\AT\ $1s^2\,2s^1\,2p^3$}                            &  & &    \\
    &                     &    &  &                                                                          &    & \multicolumn{2}{l}{\RT\ $\boldsymbol{1s^2\,2s^2\,2p^2}$}   &  & & \\
   && & \multicolumn{2}{l}{\RT\ $1s^2\,2s^1\,2p^4$}                               &         & &                                                                                         &  & & \\
  & & & & \RT\ $\boldsymbol{1s^2\,2s^2\,2p^3}$                                    &         & &                                                                                          &  & & \\
  \end{tabular}
\end{ruledtabular}
\end{table}

Extended energy-level and transition-rate calculations need to be performed to model the successive (photo-) detachment from negative ions. For inner-shell-excited atoms and ions, especially the multi-configuration Dirac-Hartree-Fock (MCDHF) method \cite{Grant2007,Fritzsche2002a} has been found to be a versatile tool to model the -- radiative and non-radiative -- decay and to describe the interplay of different atomic processes in course of their relaxation. In this work, all calculations have been performed by means of the \Jac{} toolbox, the Jena Atomic Calculator \cite{Fritzsche2019}, which has been expanded recently to follow rather long ionization pathways. Not much needs to be said about how such cascade computations are implemented in practice  \cite{Fritzsche2021,Jac-manual}, and we just summarize a few major steps that have been carried out. The numerical results of these simulations will then be discussed together with the measurements below.

The photoexcitation and ionization of an electron requires first of all insight into the $1s^{-1}$ and $(1s\, n\ell)^{-2}$ threshold energies and the associated cross sections. This refers to both the excitation and decay of the {\color{black} core-excited $1s\,2s^2\,2p^6\;^2S_{1/2}$ level of the O$^-$ ion and the core-detached O$^-$, that is  $1s\,2s^2\,2p^5\;^{1,3}P_J$ core-excited levels  of the neutral atom,} as well as of various doubly core-excited configurations. In particular, the $1s\,2s^2\,2p^4\,3p$, $1s\,2s^2\,2p^4$, and $1s\,2s\,2p^5$ configurations play a prominent role {\color{black} in  the cascade decay of $1s$ core-detached neutral oxygen} and have been included in the computations. The probabilities for different shake transitions have been determined from the orbital overlap of the $1s^2\,2s^2\,2p^5$ ground configuration and the $1s 2s^2 2p^5 + 1s\,2s^2\,2p^4\,(3p + 4p)$ configurations, and show that up to 10\% of the $2p$ electrons are \textit{shaken} to orbitals with higher $n$ or even into the continuum. A similar shake probability is expected for the $2s$ electrons, although their contributions to the photodetachment cross sections are more difficult to assess. Below, we make use of these shake probabilities for estimating the ion distributions at photon energies $\hbar\,\omega \:\gtrsim\: 450$~eV.

Apart from the excitation energies and absorption strengths for creating a $1s$ inner-shell hole, the main computational focus was placed upon the stepwise relaxation and, hence, the (relative) ion distributions that can be directly compared with the experiment. This stepwise relaxation can be formally described by an \textit{atomic cascade}, that connects ions of different charge states to each other via different (decay) processes. This relaxation proceeds until a given number of electrons is released and/or the ions cannot further decay to any lower level. Such cascades therefore require to automatically determine all single configurations that may energetically occur due to various photoemission and autoionization processes from the initially chosen hole configurations/levels, and to include a proper number of shake configurations. To support a detailed analysis of different models, we therefore distinguish in \Jac{} between (so-called) cascade \textit{computations} and \textit{simulations}, cf.~Ref.~\cite{Fritzsche2021}.

In practice, any computation starts from setting up a cascade \textit{tree}, i.e., the list of configurations (Tabs.~\ref{tab:casc1s}--\ref{tab:casc1s2s}) and associated levels that likely contribute to the relaxation of the ions. These levels are then divided into (blocks of) multiplets from which the various decay pathways can be readily derived. A pathway hereby refers to a sequence of two or more levels, which can be subsequently occupied in the course of the relaxation and whose number rapidly increases with (i) the number of open shells involved for an ion and (ii) the depth of the cascade. The major computational effort refers, however, to the \textit{representation} of the fine-structure levels for each intermediate step (block) of the cascade as well as to the computation of the transition amplitudes.

To establish a hierarchy of useful cascade models, different approaches are distinguished in \Jac{} in order to support and analyze a systematically improved representation of the fine structure and amplitudes. Unlike for Si$^-$ ions \cite{Perry-Sassmannshausen2021}, the rather simple shell structure of O$^-$ with just a $1s^2\,2s^2\,2p^5$ ground configuration enables us to apply a self-consistent treatment of all blocks of the cascades (approach B), in addition to just an \textit{averaged} single-configuration approach (approach A). Indeed, approach~A, as the simplest, neglects all configuration mixing between the bound-state levels and also restricts the computations to just  a single set of continuum orbitals for each \textit{step} of the cascade \cite{Fritzsche1992}. In contrast, approach~B applies an individual self-consistent field and set of orbitals for all blocks of the cascade, while each electron configuration in the decay tree still forms a single multiplet with a well-defined fine structure (configuration mixing) by diagonalizing the Hamiltonian matrix of this configuration. This single-configuration approach is expected to already provide a quite reasonable description of the strongest decay paths. For analyzing the ion distribution below, we calculated the stepwise decay of the three initial configurations  $1s\,2s^2\,2p^5$ (Tab.~\ref{tab:casc1s}),   $1s\,2s^2\,2p^4$ (Tab.~\ref{tab:casc1s2p}),  and $1s\,2s\,2p^5$ (Tab.~\ref{tab:casc1s2s}) and for the release of up to five electrons. These ion distributions are then combined with the shake probabilities to estimate the (relative) $m$-fold photodetachment cross sections and for comparison with the experiment. The results from these simulations are shown and discussed below in Sec.~\ref{sec:res}. Despite the improved modeling with approach B, the quite limited representation of the fine-structure levels and the need for including different shake transitions are likely the main reason for the remaining deviations of the predicted ion distributions from the measurements. In addition to the cascades that follow the creation of one $1s$ hole, we also treated in a similar manner the cascades that result from double core-hole creation. In these calculations, all together 23 configurations with 3--7 electrons and 83 cascade steps were considered, a tabulation of which similar to Tabs.~\ref{tab:casc1s}--\ref{tab:casc1s2s} would exceed one page.

\section{Results and Discussion}\label{sec:res}

\begin{figure}[b]
	\centering\includegraphics[width=0.95\columnwidth]{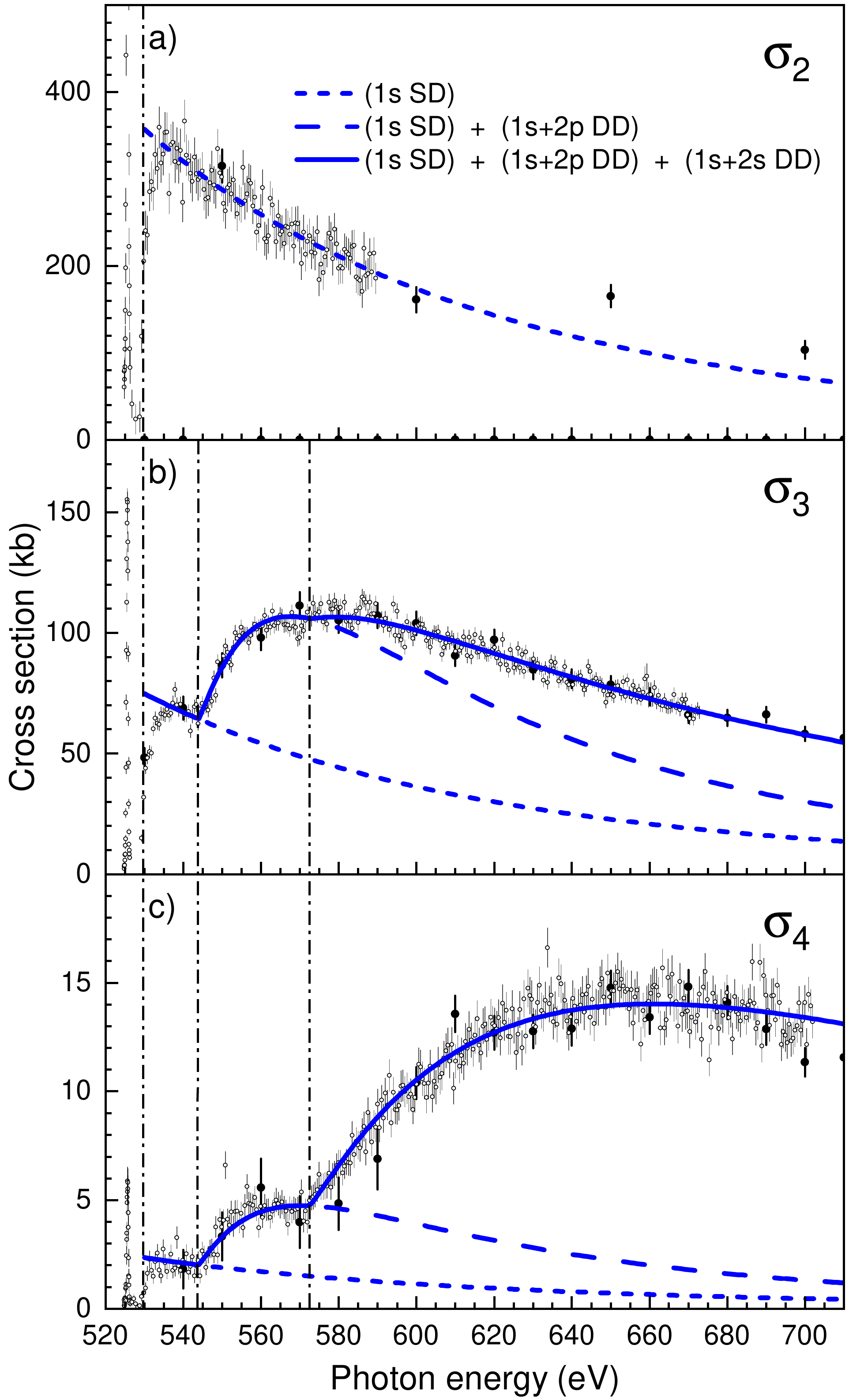}
	\caption{\label{fig:lowE}Experimental cross sections for double (a), triple (b), and quadruple (c) photodetachment of O$^-$ ions. {\color{black}The open symbols represent fine scans over a narrower energy range and the more widely spaced full symbols represent coarse scans that extend to higher energies of up to 1500~eV (see Fig.~\ref{fig:highE}).} The vertical dash-dotted lines at 529.6, 543.9, and 572.5~eV mark the thresholds for $1s$ SD, $1s$+$2p$ DD, and $1s$+$2s$ DD respectively. The full lines result from a simultaneous fit of empirical cross section formulae for these three ionization processes to the three experimental {\color{black} fine-scan} data sets (see text) with the individual contributions being represented by the dashed lines as explained by the legend in panel (a).}
\end{figure}

Figure \ref{fig:lowE} displays the measured cross sections $\sigma_2$, $\sigma_3$, and $\sigma_4$ for double, triple, and fourfold detachment of O$^-$.  The $1s\,2s^2\,2p^6\;^2S_{1/2}$ resonance at 525.6~eV can be discerned  in each of the three cross sections. Its resonance parameters  were already determined from our  previously measured high-resolution data \cite{Schippers2016a}. We used this previous analysis for  determining the experimental photon-energy spread $\Delta E$ by fitting a Voigt profile to the resonance. In the fit,  the Gaussian full width at half maximum (FWHM), $\Delta E$,  was treated as a free parameter. The fit gave rise to a spread of $\Delta E\approx 0.4$~eV.

In our previous work on multiple photodetachment of C$^-$ ions \cite{Perry-Sassmannshausen2020}, we discovered a number of previously unknown resonances, in particular, for the triple-detachment cross section. Within the limits of the present statistical uncertainties, none of the cross sections displayed in Fig.~\ref{fig:lowE} shows any signs of further resonance features in addition to the $1s\,2s^2\,2p^6\;^2S_{1/2}$ resonance at 525.6~eV. In the search for additional resonances, we have also performed scans of $\sigma_3$ with lower photon-energy spread and higher point density, but these did not result in any further resonances. We therefore conclude that the O$^-$ anion does not feature any strong excitation channels to more highly excited autoionizing levels beyond the $1s\,2s^2\,2p^6\;^2S_{1/2}$ level.

In our earlier O$^-$ experiment \cite{Schippers2016a}, the photon-energy range extended only to a few eV above the threshold for $1s$~SD at 529.6~eV. The new measurements extend to much higher energies where the cross sections $\sigma_3$ {\color{black}[Fig.~\ref{fig:lowE}(b)]} and $\sigma_4$ {\color{black}[Fig.~\ref{fig:lowE}(c)]} exhibit additional thresholds at 543.9 and 572.5~eV. According to our atomic-structure calculations (Tab.~\ref{tab:thres}), these thresholds{\color{black}, which are not discernible in $\sigma_2$ [Fig.~\ref{fig:lowE}(a)]}, correspond to the simultaneous removal  of two electrons  by one photon, i.e, to direct $1s$+$2p$~DD and to $1s$+$2s$ DD, respectively.

{\color{black} The cross section $\sigma_2$ for net double detachment does not exhibit any significant DD features, since the Auger processes, that follow the initial double core-hole creation, further increase the charge state of the intermediate O$^+$ ion (created by DD of O$^-$) with a probability of almost 100\%. Therefore, above the threshold for $1s$ SD at 529.6~eV, the cross section $\sigma_2$ is practically exclusively due to this single-detachment process. For the purpose of curve fitting, we have parameterized the cross section  for $1s$ SD  following a prescription of  Verner et al. \cite{Verner1993a}. We have determined  the according parameter values by a fit to~$\sigma_2$~\footnote{Parameters for representing the O$^-$ $1s$ SD cross section according to Eqs.~1 and 2 of Ref.~\cite{Verner1993a}: $E_\mathrm{th}=529.6$~eV, $l=0$, $E_0=188.525$~eV, $\sigma_0=16.05$~kb, $y_a=400$, $P=-0.71946$, $y_w=94.7123$}. The resulting curve  $\sigma^{(SD)}_{1s}(E)$ is displayed as the short-dashed line in Fig.~\ref{fig:lowE}(a). Above the threshold, the fit matches the experimental data. The agreement is however less convincing  at the threshold. This  can be attributed to the fact that the Verner formula, which was originally designed for photoionization cross sections of positively charged ions \cite{Verner1993a}, does not account for the more gently rising photodetachment thresholds of negative ions. Never\-theless, the fit provides a realistic extrapolation of the 1s~SD cross section towards higher energies as required below.

\begin{table*}
\caption{\label{tab:patt}Parameter values for expressing the apparent DD cross sections {\color{black} $\tilde{\sigma}_m^{(DD)} = F_m\sigma^{(DD)}(E)$ with  $\tilde{\sigma}_m^{(\mathrm{max})} = F_m\sigma^{(\mathrm{max})}$  via Eq.~\ref{eq:Pattard}  with $\alpha=1.1269$} as resulting from the fits discussed in the text.  The specific contributions of the $1s$~SD cross section to $\sigma_2$, $\sigma_3$, and $\sigma_4$ are $F_2^{(1s)}=0.8223(47)$, $F_3^{(1s)}=0.1723(26)$, and $F_4^{(1s)}=0.00540(22)$. Numbers in parentheses denote one-sigma uncertainties obtained from the fit. The values for $1s$+$1s$~DD bear additional unknown uncertainties associated with  \lq\lq{}background\rq\rq\ subtraction (see text).}
 \begin{ruledtabular}
 \begin{tabular}{cccccc}
  Process           &   $E^{(\mathrm{th})}$ (eV)  & $E^{(\mathrm{max})}$ (eV) &   $\tilde{\sigma}_3^{(\mathrm{max})}$  (kb)  &  $\tilde{\sigma}_4^{(\mathrm{max})}$  (kb)  &  $\tilde{\sigma}_5^{(\mathrm{max})}$ (kb)  \\[0.5ex]      \hline
    $1s$+$2p$ DD  &      543.92(29)                  &        575.0(10)  & 58.4(13)   & 3.25(14)  & \\
    $1s$+$2s$ DD  &       572.51(59)                  &        680.3(26)  & 28.1(38)   & 12.2(02)  & \\
    $1s$+$1s$ DD  &     1010(46)  &  1627(324)   & &   & 0.081(17) \\
 \end{tabular}
\end{ruledtabular}
\end{table*}

The $1s$ SD process also contributes to the cross sections for net triple and fourfold detachment displayed in Figs.~\ref{fig:lowE}(b) and \ref{fig:lowE}(c), respectively. However, above the thresholds for $1s$+$2p$ DD and $1s$+$2s$ DD,  these latter processes dominate the cross sections $\sigma_3$ and $\sigma_4$. The cross section for net fourfold detachment} rises from 5~kb at the $1s$+$2s$~DD thres\-hold to a maximum value of 14~kb at about 650~eV. The corresponding rise is barely visible in the cross section for triple detachment, which nevertheless exhibits a strong contribution by $1s$+$2p$ DD leading to a cross-section rise from $\sim$70~kb at the $1s$+$2p$ DD threshold to $\sim$105~kb at the $1s$+$2s$ DD threshold.

As demonstrated  earlier \citep{Mueller2018b,Perry-Sassmannshausen2020, Mueller2021},  the DD contributions to the measured cross sections can be represented as functions of the photon energy $E$ by a semi-empirical formula, that has been devised by Pattard \cite{Pattard2002}:
\begin{equation}\label{eq:Pattard}
\sigma^{(DD)}(E) =\sigma^\mathrm{(max)}x^\alpha\left( \frac{\alpha+7/2}{x\alpha+7/2}\right)^{\alpha+7/2},
\end{equation}
where  $x = ({E-E^{(\mathrm{th})}})/({E^{(\mathrm{max})}-E^{(\mathrm{th})}})$, $E^{(\mathrm{th})}$ is the threshold energy and {\color{black} $\alpha=1.1269$ is the  Wannier exponent. Its numerical value \cite{Wannier1953} is the appropriate one for the charge of the doubly detached intermediate O$^+$ ion, which results from direct double detachment of O$^-$.} Furthermore in Eq.~\ref{eq:Pattard},  $\sigma^{(\mathrm{max})}$ is the cross-section maximum, which occurs at the energy $E^{(\mathrm{max})}$. Numerical values for these parameters were obtained by fitting
\begin{eqnarray}\label{eq:fit}
\sigma_m^{(\mathrm{fit})}(E) &=& F_m^{(1s)} \sigma^{(SD)}_{1s}(E)+\\
 & & F_m^{(1s+2p)}\sigma^{(DD)}_{1s+2p}(E)+ F_m^{(1s+2s)}\sigma^{(DD)}_{1s+2s}(E)\nonumber
\end{eqnarray}
simultaneously to the three experimental cross sections $\sigma_m$ shown in Fig.~\ref{fig:lowE}. The  {\color{black}energy-independent} factor $F_m^{(1s)}$ adjusts the relative contribution of the $1s$ SD cross section $\sigma^{(SD)}_{1s}(E)$  as appropriate for each $\sigma_m$. The  factors  $F_m^{(1s+2p)}$ and  $F_m^{(1s+2s)}$ have the same role for the $1s+2p$ and $1s+2s$ DD processes, respectively. However, the fit  does not allow one to disentangle these weight factors from the {\color{black}corresponding DD} cross sections. Therefore,  Tab.~\ref{tab:patt}{\color{black}, which tabulates the fit results,  provides parameters for the \emph{apparent} DD cross sections $\tilde{\sigma}_m^{(DD)}(E) = F_m\sigma^{(DD)}(E)$ with their cross-section maxima $\tilde{\sigma}_m^{(\mathrm{max})} = F_m\sigma^{(\mathrm{max})}$ being specific for each $m$-fold detachment channel (cf.~Eq.~\ref{eq:reaction}).}

{\color{black} The $1s$+$2p$ DD and $1s$+$2s$ DD threshold energies that result from the fit (Tab.~\ref{tab:patt})} have a combined fit and systematic uncertainty of less than 1~eV. Within this uncertainty, they agree with the theoretical predictions (Tab.~\ref{tab:thres}). All cross-section maxima decrease when going from triple to fourfold detachment. For  $1s$ SD, $1s$+$2p$ DD, and  $1s$+$2s$ DD, they are reduced by factors of 0.031, 0.056, and 0.43, respectively, i.e., the relative importance of $1s$+$2s$ DD is higher for fourfold  than for triple detachment.  For triple detachment, the cross-section maximum  of $1s$+$2p$ DD is twice as large as the one of $1s$+$2s$ DD, while for fourfold detachment the latter is nearly a factor of  four larger than the former. The higher relative  yield of O$^{3+}$ ions from the $1s$+$2s$ DD process as compared to $1s$+$2p$ DD can be understood already qualitatively by comparing the respective cascade trees in Tabs.~\ref{tab:casc1s2s} and \ref{tab:casc1s2p}. In particular, the chain of the autoionizing transitions O$^+$($1s^1\,2s^1\,2p^5$) $\to$ O$^{2+}$($1s^1\,2s^2\,2p^3$) $\to$ O$^{3+}$($1s^2\,2s^2\,2p^1$), which starts with a fast Super-Coster-Kronig process, is characteristic for $1s$+$2s$ DD (Tab.~\ref{tab:casc1s2s}).

\begin{figure}
	\includegraphics[width=\columnwidth]{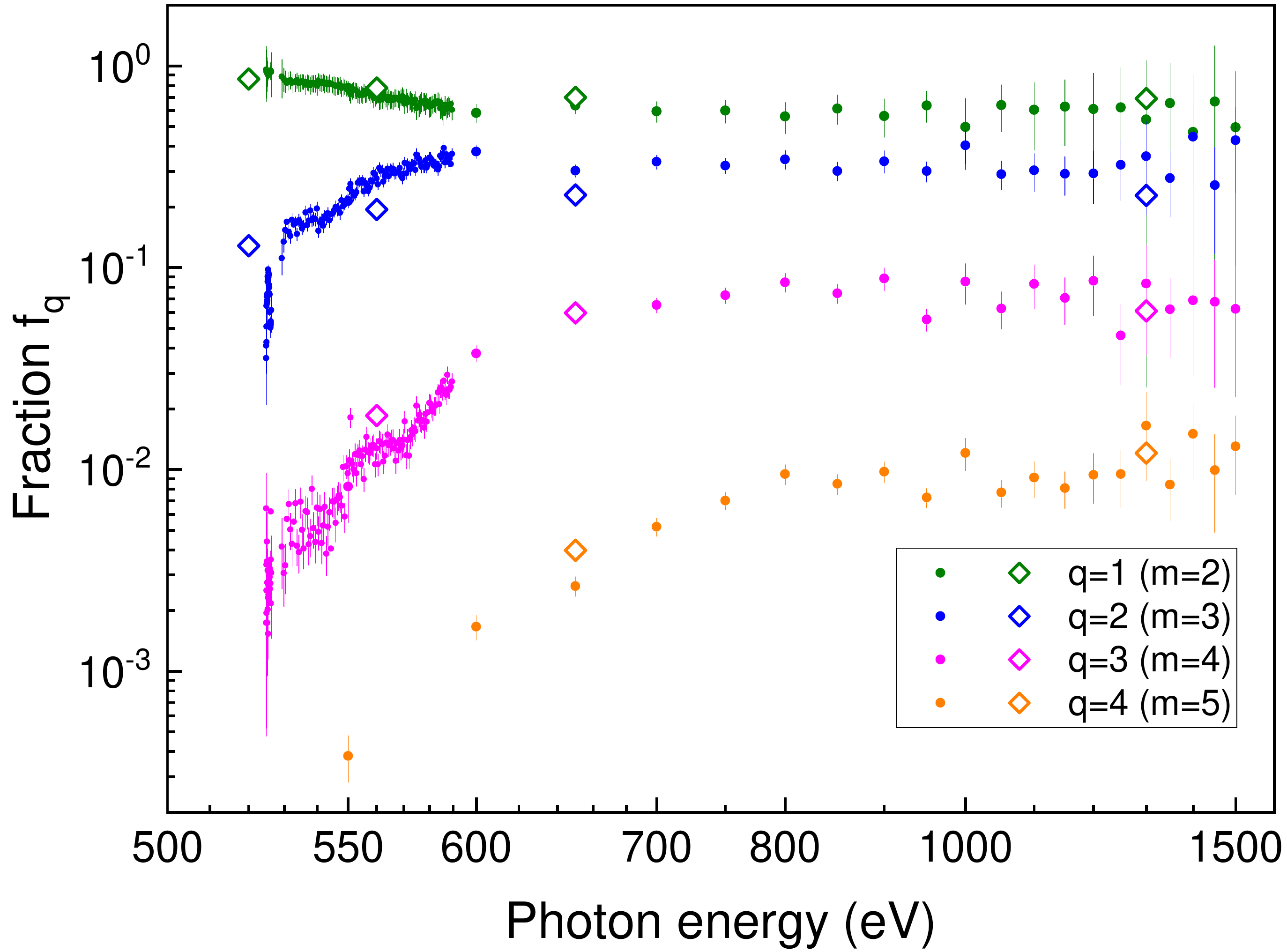}
	\caption{\label{fig:frac}Experimental (full circles) and theoretical (open diamonds) product-ion charge fractions $f_q$ (Eq.~\ref{eq:frac}). The photon-energy ($E_\mathrm{ph}$) axis is compressed towards higher energies using $x=\log(E_\mathrm{ph}/\mathrm{eV}-400)$ for the abscissa.}
\end{figure}

\begin{table}
\caption{\label{tab:casc}Theoretical product-ion charge-state distributions resulting from the present cascade calculations for $1s$ SD ($F_m^{(1s)}$), $1s$+$2p$ DD  ($F_m^{(1s+2p)}$),  $1s$+$2s$ DD  ($F_m^{(1s+2s)}$),  and $1s$+$1s$ DD  ($F_m^{(1s+1s)}$). The quantity $q$ denotes the product-ion charge state which equals $m-1$ (Eq.~\ref{eq:reaction}).}
 \begin{ruledtabular}
 \begin{tabular}{cccccc}
$q$ &   $m$        &   $F_m^{(1s)} $  & $F_m^{(1s+2p)}$ &   $F_m^{(1s+2s)}$  & $F_m^{(1s+1s)}$  \\[0.5ex]
   \hline
  0  &  1 & 0.011 & 0.0       & 0.0      & 0.0  \\
  1 &  2 & 0.861 & 0.029  &  0.069  & 0.001  \\
  2 &  3 & 0.128 & 0.786  &  0.478  & 0.069 \\
  3  & 4 & 0.0     &  0.185  & 0.413  & 0.120 \\
  4  & 5 & 0.0    &  0.0       & 0.040  & 0.810 \\
  5 &6 & 0.0    & 0.0       & 0.0       & 0.0
 \end{tabular}
\end{ruledtabular}
\end{table}

Our fine-structure-resolved cascade computations, which in addition to single-step autoionizing transitions also consider competing radiative transitions, can be compared quantitatively with the experimental findings. Figure~\ref{fig:frac} compares experimental and theoretical product-ion charge fractions
\begin{equation}\label{eq:frac}
f_q =  \frac{\sigma_{q+1}}{\sigma_\Sigma},
\end{equation}
where $q=m-1$ is the product-ion  charge state resulting from net $m$-fold detachment and
\begin{equation}\label{eq:sum}
\sigma_\Sigma = \sum_{m=2}^5\sigma_m = \sum_{q=1}^4\sigma_{q+1}.
\end{equation}

Theoretical $f_q$ values have been computed for the photon energies  520, 560, 650, and 1300~eV (Fig.~\ref{fig:frac}), which are above the thresholds for $1s$ SD, $1s$+$2p$ DD, $1s$+$2s$ DD,  and $1s$+$1s$ DD, respectively.  In these computations, the  $F_m$ values  which result from our cascade calculations and which are provided in Tab.~\ref{tab:casc} have been weighted with the relative cross sections for the individual SD and DD processes as discussed in the following.

At 520~eV,  a $1s$ hole can be created only  via $1s$ SD. Accordingly, $f_q(520\mathrm{~eV})=F_m^{(1s)}$. The subsequent single-step cascade tree produces O$^{2+}$ as the highest-charged product ion (Tab.~\ref{tab:casc1s}). The theoretical O$^+$ and O$^{2+}$ fractions of 86\% and 13\% (Tab.~\ref{tab:casc}) agree well with the  corresponding experimental $F_m^{(1s)}$ values (caption of Tab.~\ref{tab:patt}). A fraction of 1\% is theoretically predicted for neutral oxygen, which cannot be observed in the present experimental configuration.  The small experimentally observed O$^{3+}$ fraction can only be explained if higher-order processes are taken into account such as multiple Auger processes (see, e.g., \cite{Mueller2015a,Mueller2021b}).  Such processes were, however, not considered in the present cascade calculations.

At 560~eV, the $1s$+$2p$ DD process is energetically allowed in addition to  $1s$ SD. Now the maximum calculated charge state, that can be reached by single-step deexcitation cascades, is 3+ (Tab.~\ref{tab:casc1s2p}). In the absence of a rigorous theoretical treatment of direct double-detachment processes, we  just assumed in our calculations that  the $1s$+$2p$ DD process contributes by 10\% to the total detachment cross section, which  is of the same order of magnitude as the shake probabilities mentioned in Sec.~\ref{sec:theo}. Accordingly, $f_q(560\mathrm{~eV})= 0.9F_m^{(1s)}+0.1F_m^{(1s+2p)}$. The resulting fractions of net double, triple, and fourfold detachment agree surprisingly well with the experimental findings considering the coarseness of this approach. The same holds at 650~eV, which is above the threshold for $1s$+$2s$ DD and where our cascade calculations also predict a nonzero contribution by net fivefold detachment.  In analogy to the $1s$+$2p$ DD process,  $1s$+$2s$ DD also received a weight of 10\%  such that  $f_q(650\mathrm{~eV})= 0.8F_m^{(1s)}+0.1F_m^{(1s+2p)}+0.1F_m^{(1s+2s)}$. At 1300 eV, i.e., above the threshold for $1s$+1$s$ DD, a 1\% contribution from this latter process has been assumed in addition and, correspondingly, $f_q(1300\mathrm{~eV})= 0.79F_m^{(1s)}+0.1F_m^{(1s+2p)}+0.1F_m^{(1s+2s)}+0.01F_m^{(1s+1s)}$.

\begin{figure}
	\includegraphics[width=\columnwidth]{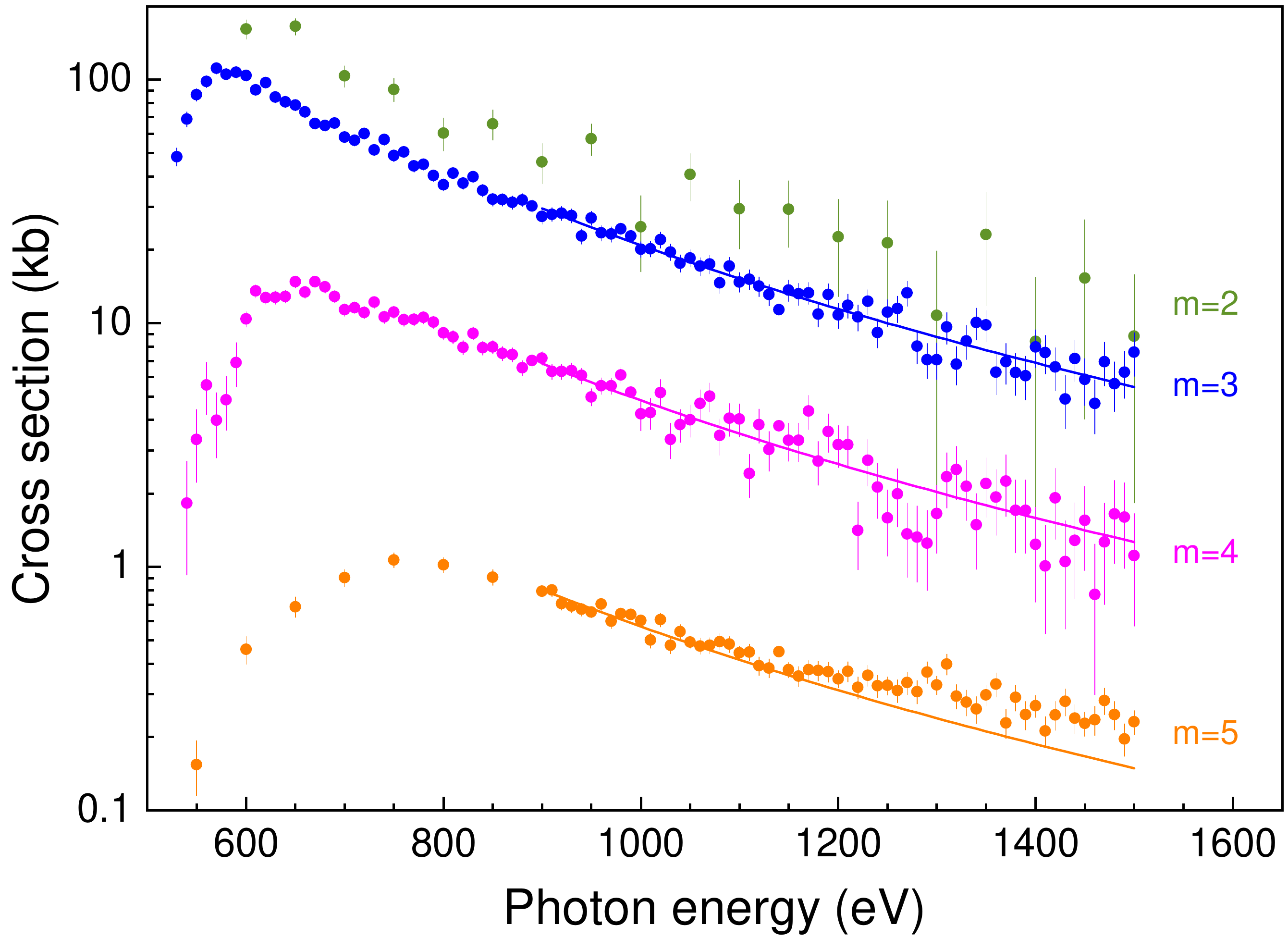}
	\caption{\label{fig:highE}Results from coarse photon-energy scans over a wide range.  The differently colored data points represent the experimental cross sections $\sigma_m$ for multiple ($m$-fold, Eq.~\ref{eq:reaction}) detachment  of O$^-$ ions for  $m$=2 (olive), $m$=3 (blue), $m$=4 (magenta), and $m$=5 (orange).  The error bars account only for the statistical experimental uncertainties. The full lines represent a power-law with the exponent $-3.3\pm0.1$ as obtained from a fit to $\sigma_3$ (see text).}
\end{figure}

In our measurements, we scrutinized the weak O$^{4+}$ product-ion channel  in the search for a signature of double $K$-hole formation via the $1s$+$1s$ DD process. According to our calculations, the threshold for direct double $K$-shell detachment  occurs at 1160~eV (Tab.~\ref{tab:thres}). Around this energy, the cross section $\sigma_5$ exhibits a noticeable rise which is not present in $\sigma_3$ and $\sigma_4$ (Fig.~\ref{fig:highE}). These latter two cross sections  show identical high-energy behaviors, which can be described by a power law with an exponent of  $-3.3$. This number was obtained from a fit to $\sigma_3$, and the resulting fit curve was then scaled to $\sigma_4$ and $\sigma_5$ by multiplication with appropriate factors (full lines in Fig.~\ref{fig:highE}).

\begin{figure}
	\includegraphics[width=\columnwidth]{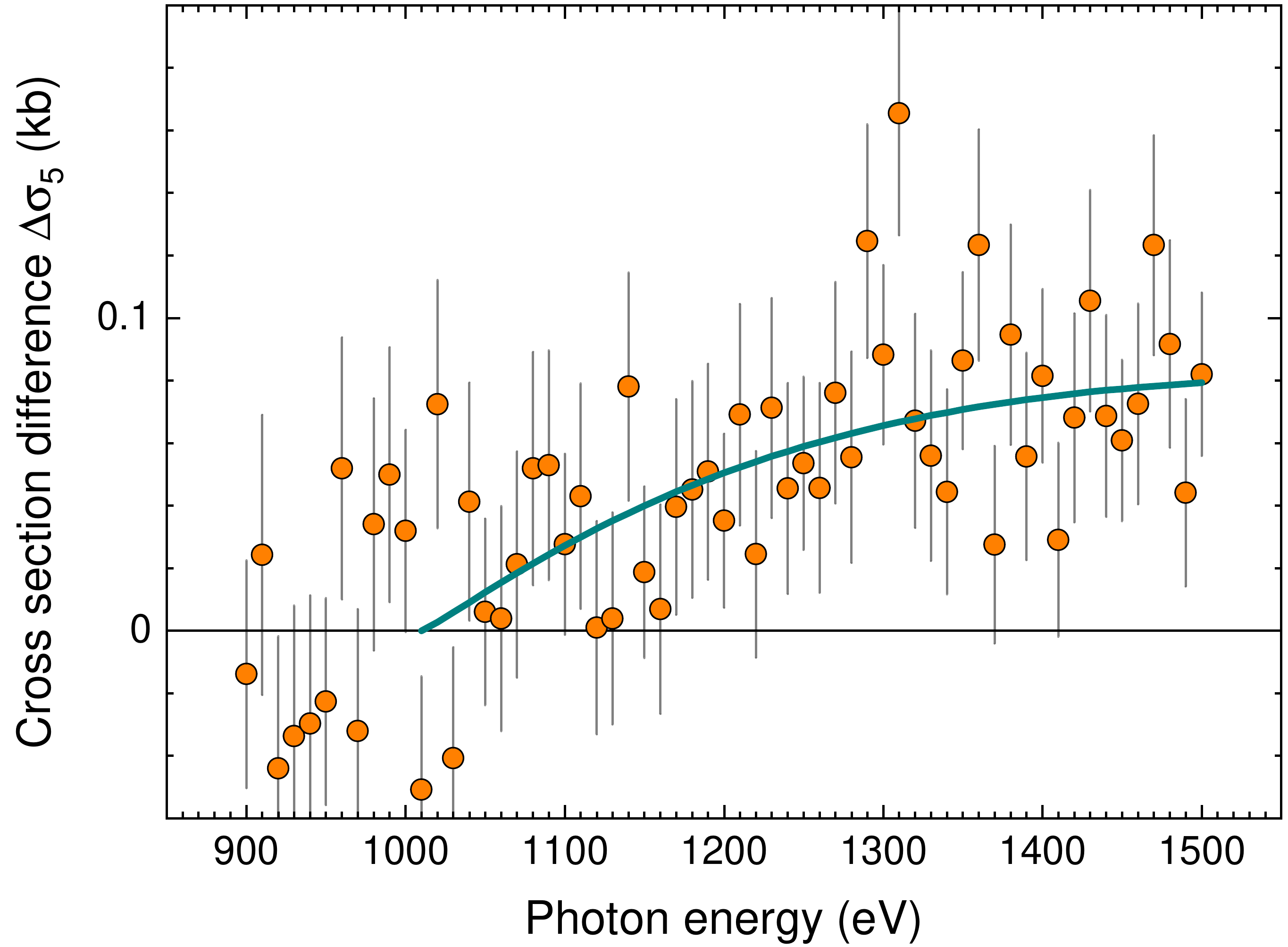}
	\caption{\label{fig:twoK}Cross-section difference resulting from the subtraction of the scaled power-law fit from the measured cross section $\sigma_5$ (see Fig.~\ref{fig:highE}). The full line results from a fit  of the Pattard formula (Eq.~\ref{eq:Pattard}) to the data points. The fit results are listed in Tab.~\ref{tab:patt}.}
\end{figure}

The data points in Fig.~\ref{fig:twoK} were obtained by subtracting the appropriately scaled  power law from the measured cross section $\sigma_5$. This cross-section difference $\Delta\sigma_5$ is zero up to about 1000~eV from where about it rises towards higher energies. A fit of Eq.~\ref{eq:Pattard} to the data points yields a threshold value of $1010\pm46$~eV (Tab.~\ref{tab:patt}) which is somewhat lower than our calculated value. Conceding  an additional (unknown) uncertainty related to the subtraction of the power-law curve,  it seems nevertheless plausible that the cross-section difference depicted in Fig.~\ref{fig:twoK} is indeed due to $1s$+$1s$ DD.

The fit of the Pattard formula (Eq.~\ref{eq:Pattard}) yields a maximum value of the $1s$+$1s$ DD cross section of  $81\pm17$~b. This is significantly smaller than the maximum cross section of 3~kb which we have found earlier for fivefold detachment of C$^-$ \cite{Perry-Sassmannshausen2020}. This can be (at least partly) attributed to the general {\color{black}strong} decrease of {\color{black}double-}photoionization cross sections with increasing nuclear charge~{\color{black}\cite{Hoszowska2009}}. Moreover, it cannot be excluded that $1s$+$1s$ DD also contributes to the production of final ion charge states other than O$^{4+}$. Our cascade calculations result in a product charge-state distribution that consists of 7\% O$^{2+}$, 12\% O$^{3+}$, 81\% O$^{4+}$, and 0\%  O$^{5+}$ ions ($F_m^{(1s+1s)}$ in Tab.~\ref{tab:casc}). At the present level of statistical uncertainty, a $\sim$10\%  contribution of $1s$+$1s$ DD to $\sigma_4$ can  remain unnoticed considering the scatter of the experimental $\sigma_4$ data points above 1000~eV (Fig.~\ref{fig:highE}). We also scrutinized the O$^{5+}$ channel, but the count rate was too low for a measurement of $\sigma_6$ with satisfying statistical uncertainty in a reasonable amount of time.

It should be noted that the measurement of the small cross section $\sigma_5$, which is  in the sub-kilobarn range, is particularly challenging. In addition, at photon energies above 1000~eV, the photon flux at the PETRA\,III photon beamline P04 decreases  rapidly as the photon energy increases, so that it becomes difficult to measure small cross sections with sufficient statistical accuracy when going to higher energies. In the present experiment, the lowest O$^{4+}$ count rate was only 0.2~Hz. Nevertheless, this is still an order of magnitude larger than the dark-count rate ($\sim$0.02~Hz, see Sec.~\ref{sec:exp}) of our single-particle detector. This rules out that the rise of $\sigma_5$ beyond $\sim$1100~eV is an artifact of our product-ion detection scheme. In this context we mention our recent measurement of the similarly small cross section for sixfold photoionization of Ar$^+$ ions \cite{Mueller2021b}.

\section{Summary and Conclusions}\label{sec:conc}

Employing the photon-ion merged-beams technique at a high-flux beamline of one of the world's brightest synchrotron-radiation facilities, we were able to measure cross sections for the single-photon multiple inner-shell photodetachment of oxygen anions over a wide energy range extending to well beyond the threshold for double $K$-shell detachment. Owing to a high product-ion selectivity and a near 100\% product-ion detection efficiency, we were able to measure cross sections for net double, triple, fourfold, and fivefold ionization. These turned out to be dominated by direct double-detachment processes where one photon simultaneously ejects two electrons.  As in our previous studies on F$^-$ \cite{Mueller2018b} and C$^-$ \cite{Perry-Sassmannshausen2020}, the experimental cross-section shapes are very well described by the empirical scaling formula suggested by Pattard \cite{Pattard2002}. {\color{black} In future experiments, more detailed information might be obtained from electron spectroscopy. Although this is rather demanding, mainly because of the low densities of ionic targets, first steps in this direction have already been taken \cite{Domesle2010,Harbo2012,Bizau2016}.}

We identified the various single and double detachment thresholds by large-scale atomic-structure calculations which also account for the complex deexcitation cascades that set in after the initial core-hole creation. The theoretical results for the product charge-state distributions agree well with the experimental findings despite of the fact that only single Auger and radiative dipole transitions were  considered in the computation of the deexcitation cascades. The calculations do not predict any yield of O$^{4+}$ product ions below the threshold for $1s$+$1s$ double ionization although a small experimental cross for fivefold detachment could be measured. This suggests that many-electron processes such as detachment accompanied by shake-up or  double-Auger processes are decisive for the production of O$^{4+}$ ions. In the future, we will incorporate such processes into our systematic approach for the computation of deexcitation cascades~\cite{Fritzsche2021}. Even more theoretical development work is required for  a  more quantitative understanding of the various double-ionization processes at play.\\

\begin{acknowledgments}
We acknowledge DESY (Hamburg, Germany), a member of the Helmholtz Association HGF, for the provision of experimental facilities. Parts of this research were carried out at PETRA\,III and we would like to thank  Kai Bagschik, Frank Scholz, J\"orn Seltmann, and Moritz Hoesch for assistance in using beamline P04. We are grateful for support from Bundesministerium f\"{u}r Bildung und Forschung within the \lq\lq{}Verbundforschung\rq\rq\ funding scheme (grant nos.\ 05K19GU3 and 05K19RG3) and from Deutsche Forschungsgemeinschaft (DFG, project nos.\  389115454 and SFB925/A3).
\end{acknowledgments}

%\bibliography{/tex/AMPstefan}
%\end{document}

%apsrev4-2.bst 2019-01-14 (MD) hand-edited version of apsrev4-1.bst
%Control: key (0)
%Control: author (8) initials jnrlst
%Control: editor formatted (1) identically to author
%Control: production of article title (0) allowed
%Control: page (0) single
%Control: year (1) truncated
%Control: production of eprint (0) enabled
%

\end{document}